\documentstyle[11pt]{article}
\date{}
\begin{document}
\sloppy
\author{V. Majern\'{\i}k$^{* \dag}$\\
${}^{*}$ Institute of Mathematics, Slovak
Academy of Sciences, \\ Bratislava, \v Stef\'anikova  47,
 Slovak Republic\\and\\
$^{\dag}$Department of Theoretical Physics,
Palack\'y University,\\ T\v r. 17. Listopadu 50, CZ-772 07 Olomouc,
Czech Republic}
\title{Cosmological model with $\Omega_M$-dependent cosmological
constant}
\maketitle
\begin{abstract}
The idea here is to set the cosmical constant $\lambda$ proportional to
the scalar of the stress-energy tensor of the ordinary matter.
We investigate the evolution of the scale factor in a
cosmological model in which the cosmological constant is proportional to
 the scalar of the stress-energy tensor.
\end{abstract}

\section{Introduction}

The observational view of the universe has drastically changed during
the last ten years. New observation suggests a
universe that is leigh-weight, is accelerating, and is flat
\cite{PER} \cite{pe} \cite{BO}.
One way to account for cosmic acceleration is the
introduction a new type
of energy, the so-called {\it quintessence} ("dark energy"), a dynamical,
spatially inhomogeneous form of energy with negative pressure \cite{SH}.
A common example is the energy of a slowly evolving scalar field with positive
potential energy, similar to the inflation field in the inflation
cosmology.
The quintessence cosmological scenario (QCDM) is a
spatially flat FRW space-
time dominated by the radiation at early time,
and cold dark matter (CDM) and quintessence (Q) at later time.
A series of papers of Steinhardt et al. is devoted to
the various quintessence cosmological models \cite{ST}
(a number of follow-up studies are underway). The
quintessence is supposed to obey an equation of state of the form
\begin{equation} \label{1}
p_Qc^{-2}=w_Q\varrho_Q,\qquad
-1<w_Q<0.
\end{equation}
In many models $w_Q$ can vary over time. For the vacuum energy
(static cosmological
constant), it holds $w_Q=-1$ and $\dot w_Q=0$.

In what follows we present
a variant of the quintessence cosmological scenario in which the content of
black energy is given by the cosmological constant.
The possible existence of very small but non-zero cosmological constant
revives in these days due to new observation in cosmology.
In the absence of a symmetry in nature which would set the value of
$\lambda$ to precisely zero, one is forced to either set $\lambda\neq 0$
by hand, or else look for mechanisms that can generate
$\lambda=\lambda_{obs}>0$, where $\Lambda\approx 10^{29}g cm^{-3}$
is the value of the $\Lambda$-term inferred from recent supernovae
observation. There are several  mechanism which could, in principle, give
rise
either to time independent constant, or else a time dependent
$\Lambda$-term.
Models
with a fixed $\Lambda$ run into fine-tuning problems since the ratio of the
energy density in $\Lambda$ to that of matter/radiation must be tuned to
better than one part in $10^{60}$ during the early universe in order that
$\Lambda \approx \rho_{matter}$ today. Scalar field models
considerably alleviate this problem
though some fine-tuning does remain in determining the 'correct
choose' of parameters in the scalar field potential.

Due to this fact, there are many
phenomenically ansatzes for the
cosmological constant, e.g. the different built-in cosmological
constants
(for a detailed analysis of these models see \cite{ARB}) which are
more or less justified by the physical arguments. We remark that observational
data indicate that $\lambda\approx 10^{-55} cm^{-2}$ while
particle physics prediction
for $\lambda$ is greater than this value by factor of order $10^{120}$.
This discrepancy is known as the {\it cosmological constant problem}.
The vacuum energy assigned to $\lambda$
appears very tiny but not zero. However, there is no
really compelling dynamical explanation for the smallness of the vacuum
energy at the moment \cite{Is}
(simple quantum-mechanical calculations yield
the vacuum energy much larger \cite{WW}).
The quintessence eventually modelled by a positive non-zero
cosmological constant helps
overcome the age problem, connected on the one side
with the hight estimates of the Hubble parameter and with the age of globular
clusters on the other side.
To explain this apparent discrepancy the point of view has often been adopted
which allows the cosmological constant to vary in time. The idea is
that during the
evolution of universe the "black" energy linked with cosmological constant
decays into the particles causing its decrease.

It is well-known that the Einstein field equations with
a non-zero $\lambda$ can be rearranged so
that their right-hand sides consist of two terms:
the stress-energy tensor of the ordinary
matter and an additional tensor
\begin{equation}  \label{2}
T^{(\nu)}_{ij}=\left (\frac{c^4\lambda}{8\pi G}\right)
g_{ij}=\Lambda g_{ij}.
\end{equation}
In common discussions, $\Lambda$ is identified with vacuum energy
because this quantity satisfies the requirements asked from $\Lambda$, i.e.
(i) it should have the dimension of energy density, and (ii) it
should be invariant under Lorentz transformation. The second property
is not satisfied for arbitrary systems, e.g. material systems and
radiation. Gliner \cite{G} has shown that the
energy density of vacuum represents a scalar function of the
four-dimensional space-time coordinates so that it satisfies both above
requirements. This is why  $\Lambda$ is commonly {\it identified}
with the vacuum energy.

However, there may be generally other quantities
satisfying also the above requirements.
Instead of identifying $\Lambda$ with the vacuum energy
we have identified $\Lambda$ in \cite{MAL}
with the stress-energy scalar $T=T^i_i$ a scalar
which
arises by the contraction of the
stress-energy tensor of the ordinary matter $T^{j}_i$.
This quantity {\it likewise} satisfies both above requirements, i.e.,
it is Lorentz invariant and has the dimension of the energy density.
Hence, we make the ansatz
\begin{equation} \label{3}
\Lambda_A=\frac{c^4 \lambda_A} {8\pi G}= \kappa T^i_i= \kappa T
\end{equation}
or
\begin{equation} \label{4}
\lambda_A=\frac{8\pi G\kappa T}{c^4},
\end{equation}
where $\kappa$ is a dimensionless constant to be determined.
$\Lambda_A$ is a dynamical quantity, changing over time, representing, in the
quintessence theory, the quintessence component.
In contrast with some other cosmological models, we
suppose that the universe consists of a mixture of the {\it ordinary}
mass-energy
and the quintessence component functionally linked with $T$ via the
cosmological constant $\lambda_A$.
We note that there are similar attempts to identify $\lambda$ with
the Ricci scalar (see \cite{AL}).

We describe a cosmological model in which we consider (3) as a
phenomenological ansatz for the cosmological constant.
Assuming the
flatness of space the constant $\kappa$ is uniquely given.
This model of
the universe we confront with the observation and find that it is in
concord with the data.
The word "phenomenically" means that no attempt to derive these models
from the underlying quantum field theory is being made.
Historically, an array of the  phenomenological $\Lambda$-models were
proposed since
1986. These model may be classified into two groups: (i) kinetical
models where $\Lambda$ is simple assumed to be function of either the
cosmological time $t$, the scale factor $a(t)$, etc., of the FRW
cosmological model and (ii) field-theoretical models. Here the
$\Lambda$-term
is assumed to be new physical classical field with some
phenomenological Hamiltonian. The phenomenological model introduced here
does not belong to any of these classes since $T$ is not a kinetical
quantity; rather it can be considered as a model
with possible field-theoretical background.

\section{Friedmann's model with a $\Omega_M$-dependent cosmological
constant}
The standard Einstein field equations (see, e.g. \cite{W})
can be written in the form
\begin{equation}
\label{6}
R_{ij}-g_{ij}(1/2)R=\frac{8\pi G}{c^4}(T^{(m)}_{ij}+T^{(v)}_{ij}),
\end{equation}
where $T_{ij}^{(m)}$ is the energy-momentum tensor for the perfect fluid
[23]
\begin{equation} \label{a}
T_{ij}^{(m)} =(\rho + p/c^2)u_iu_j -p g_{ij}
\end{equation}
and $$T^{(v)}_{ij}=g_{ij}\Lambda\quad\Lambda={\lambda c^4\over
8\pi G}.\quad\eqno(5a)$$
Putting $\Lambda=\Lambda_A=\kappa T$ we have
$T^{(v)}_{ij}=g_{ij}\kappa T.$
Inserting Eqs.(5a) and (5) into Eq.(\ref{6}) we have
\begin{equation}\label{b}
R_{ij}-g_{ij}(1/2)R = \frac{8\pi
G}{c^4}\left [(\rho + p/c^2)u_iu_j -(p-\kappa T)g_{ij}\right].
\label{7}
\end{equation}
The stress-energy tensor of the cosmic medium $T^{i}_{j}$ in the
everywhere local rest frame has only four non-zero components
$T^0_{0}=\varrho c^2,  T^1_1=T^2_{2}=T^3_3=-p$ \cite{UL}. Therefore,
\begin{equation} \label{T}
T=\varrho c^2-3p.
\end{equation}
Inserting Eq.(8) into Eq.(7) we get
\begin{equation}
(T^{(m)}_{ij}+T^{(v)}_{ij})=(\rho + p/c^2)u_iu_j
-[p(1+3\kappa)-\kappa \rho c^2]g_{ij},
\end{equation}
This is the energy-momentum tensor for a perfect fluid with effective
density $\hat \rho$ and pressure $\hat p$.
\begin{equation}
T^{(m)}_{ij}+T^{(v)}_{ij}=(\hat \rho+\hat p/c^2)u_iu_j
-\hat p g_{ij}.
\end{equation}
The quantities $\hat \rho$ and $\hat p$ can be determined given the
equation of state $p=w\rho c^2$.

Our next main concern will be to find the evolution of scale factor
for Friedmann's equation (7) in
the radiation-dominated and matter-dominated eras.
In a homogeneous and isotropic universe characterized by the
Friedmann-Robertson-Walker line element the Einstein equations with
matter in the form of a perfect fluid
and non-zero cosmical term $\lambda$ acquire the following forms
\begin{equation} \label{8}
3{\dot R(t)^2\over R(t)^2}= 8\pi G\rho +\lambda
c^2-3\frac{kc^2}{R^2(t)},
\end{equation}
and
\begin{equation}                       \label{9}
\ddot R(t)= \frac{4\pi G}{3}(-\rho - 3p/c^2)+\frac{\lambda c^2}{3}R(t),
\end{equation}
where $R(t)$ is the time-dependent scale factor.

A quantitative analysis of solutions to Eqs.(11) and
(\ref{9}) can be
gained by eliminating $\rho $ in these equations and combining them into
a single equation for the evolution of the scale factor in the presence
of a $\lambda$-term \cite{XI}
\begin{equation}    \label{11}
\frac{2\ddot R}{R}+(1+3w)(\frac{\dot R^2}{R^2}+\frac{kc^{2}}{R^2})
-(1+w)\lambda
c^2=0,
\end{equation}
Here, we set the equation of state in form $p=\omega \varrho c^2$.

To determine $\Lambda_A$ which is to be inserted in
Eqs.(8) and (9) we have to specify $\kappa$.
The dimensionless constant $\kappa$  we determine by assuming that
the universe is flat, i.e., $\Omega_{tot}=1$ which is consistent with
the inflationary cosmology
density $(\Omega_{tot}=1)$ and conformed
by the measurement of the
cosmic microwave background anisotropy [18]. Since $\Omega_M<1$ we
suppose that the remaining energy of cosmological constant required to
produce a geometrical flat
universe is given by the equation
$$\Omega_M+\Omega_{\Lambda}=\Omega_M+\kappa \Omega_M=1.$$
This gives
$$\kappa=\frac{1}{\Omega_M}-1. \quad\eqno(10)$$
Inserting Eq.(10) into Eq.(3) we get
\begin{equation} \label{S}
\Lambda_A=\left (\frac{1}{\Omega_M} -1\right ) T=
\left (\frac{1}{\Omega_M} -1\right )(\rho c^2-3p).
\end{equation}

In the radiation-dominated  $p=\rho c^2/3$, hence $\Lambda_M=0$, i.e.
the cosmological evolution in this era is described by the standard FRW
model with zero cosmological constant.
All processes
which took place during this era (e.g., nucleosynthesis etc.)
are described by the
standard model.
However, in the matter-dominated era with $p=0$ and we get the following
effective density and pressure
$$\hat \rho =\rho +\kappa \rho ,\qquad \hat p=-\kappa \rho
c^2.$$
Since in the radiation-dominated era is described by standard model we
will not further deal with it, instead we will
investigation the evolution of the scale factor in dependence on
$\Omega$ in the matter-dominated era.

\section{Matter-dominated epoch}
While in the pressure-dominated universe the effect of the cosmological
constant on the evolution
of the scale factor is zero, in the matter-dominated era it affects this
evolution considerably.
Inserting $T=\rho_M c^2$ into the equation for $\Lambda_A$ yields
\begin{equation}  \label{13}
\Lambda_A=\kappa \varrho_Mc^2 =\left (\frac{1}{\Omega_M}-1\right
)\rho_M c^2
=(\varrho_{crit}-\varrho_M.)c^2=\varrho_{crit}(1-\Omega_M)c^2,
\end{equation}
We obtain the critical density $\rho_{crit}$
by means of Eqs.(\ref{13}) and Eq.(\ref{8})
\begin{equation} \label{O}
 8\pi G\rho_{crit}=\frac{3\dot R^2}{R^2}.
\end{equation}
Inserting Eq.(\ref{O}) into Eq.(14) we get immediately
the equation for the evolution of $R(t)$ for the matter dominated era
\begin{equation} \label{14}
\ddot R(t)= \left (1-\frac{3}{2}\Omega_M(t) \right) \frac{(\dot
R(t))^2}{R(t)}.
\end{equation}
Its exact solution can be found for an arbitrary time function
$\Omega_M(t)$. With the ansatz $R=\exp(y)$, we have
$$ \dot R=\dot y\exp(y), \qquad \ddot R= (\ddot y+(\dot
y)^2)\exp(y)$$
which inserting into Eq.(\ref{14}) yields
$$-(2/3)\Omega_M(t)(\dot y)^2=\ddot y.$$
By putting $\dot y =q$, this equation becomes
the form
$$-(2/3)\Omega_M(t)=\frac{\dot q}{q^2},$$
the solution to which is
$$q=\frac{1}{\int (2/3)\Omega_M(t) dt +C_1}.$$
Since $\dot y =q$ we have
$$y=\int \left(\frac{1}{\int (2/3)\Omega_M(t) dt+C_1}\right) dt +C_2.$$
With $y(t)$, the general solution of Eq.(\ref{14}) is
\begin{equation} \label{15}
R(t)=\exp \int\left (\frac{1}{\int (2/3)\Omega_M(t) dt+C_1}\right) dt
+C_2,
\end{equation}
where $C_1$ and $C_2$ are the integration constants.

In what follows we assume that $\Omega_M$ does not change
during the matter-dominated era, therefore, the solution of Eq.(18) is
$$R(t)=\left (\frac {3}{2}\right)^{\frac{2}{3\Omega_M}}(\Omega_M
C_1t+\Omega_M C_1C_2)^{\frac{2}{3\Omega_M}}.\quad\eqno(19)$$

To go further we have to specify $\Omega_M$  and the boundary conditions
for the differential equation (18). For $\Omega_M$ we take its
observable value.
There is growing observational evidence
that the total matter of the universe is significantly less than the
critical density. Several authors \cite{OS} \cite {KT}
\cite {TW} have found that the best and simplest fit is provide by
($h=0.65\pm 0.15$)
$$\Omega_M = \Omega_{CDM}+ \Omega_{baryon}\approx [0.30\pm 0.10]
+[0,04+\pm 0.01].$$
As the boundary condition we set $R(t=0)= 0$. Inserting $\Omega_M=1/3$
into Eq.(19) and respecting the the previous boundary condition,
 the evolution factor $R(t)$ take the form
$$R(t)_{(\Omega_M=1/3)}=\frac{C_1 t^2}{4}.\quad\eqno(20)$$

The time-dependence of the scale factor (20)
implies a model of the universe with the following properties:\\
(i) The Hubble parameter $H=\dot R/R=2/t$, i.e.
the age of this universe $t_0$ is approximately $2.10^{10}$ yr.
In the cosmological model with $\Lambda_A$, the universe is old
enough for the evolution of globular clusters.\\
(ii) The decelerator $q_0$ is an important parameter of any model of the
universe. It probes the equation of state of matter and the cosmological
density parameter. In our model, it takes the value $q_0=-1/2$, i.e. the
universe is accelerated in concord with the recent observation.\\
(iv) The cosmical constant $\Lambda_A$ is time-dependent
$\lambda_A=8/(c^2t^2)$. It is interesting that $\lambda\propto t^{-2}$
was phenomenologicaly set by several authors [18]-[22].\\
(v) In the considered model the universe is causally connected. The
proper distance $L(t)$ to the horizon, which is the linear
extent of the causally connected domain, diverges
$$L(t)=R(t)\int_0^t\frac{d\tau}{R(\tau)}=C_2t^2[-(\frac{2}{\tau})|^{t}_0]
=-\infty, $$
In \cite{PS} is shown that the only way to make the whole of the
observable universe causally connected is to have a model with infinite $L(t)$
for all $t>0$, i.e. in our model
the whole  observable universe is causally connected.
We remark that for $\Omega_M=1$,
$R(t)\propto t^{2/3}$, i.e. the evolution law of
$R(t)$ is in a pressure-free medium, identical with that of Standard
Cosmology.

As $\Omega_M$ decreases, $R(t)$ passes smoothly to the form
$$R(t)=\exp (C_1(t-C_2)).\quad\eqno(21)$$
It is tempting to choose for the early universe
$\Omega_M=0$, i.e. to suppose that the universe started in a massless
state and its mass content was created later through
the decay of the cosmical term.
Under this assumption we have
\begin{equation} \label{X}
R(t)=\exp(C_1(t-C_2))=R_0\exp (C_1t),\qquad C_1=\frac{1}{t_0}.
\end{equation}
The natural measures for length and time in cosmology is the Planck
length and time, i.e., $l_p=(Gh/c^3)^{1/2}=4.3.10^{-35} \rm
{m}$ and $t_p=(Gh/c^5)^{1/2}=1.34.10^{-43}\rm {s},$ respectively.
It seem to be reasonable to assume that at the very beginning of the cosmic
evolution the radius of the universe was of the order of the Planck
length, therefore we put in Eq.(19) the integration constant $R_0$ and
$C_1$ equal to $l_p$ and $1/t_p$, respectively.
Then, we get for
the initial radius and the velocity the values
$$R(0)=l_p=4.3.10^{-35}m.\qquad\dot R(0)=\frac{l_p}{t_p}=c=3.10^{8}
ms^{-1},$$
respectively. The most interesting feature of this universe is its
inflationary character.

In order to vanish
the covariant divergence of the right-hand side of Eq.(6)
the matter is created
along with energy and momentum.
Therefore, the cosmological constant $\lambda_A$ is decaying
and transforming
its energy into particles
and/or radiation. Observationally, such an effect can, in principle,
be tested: in the case of dissipative, baryon number conserving
decay of a $\Lambda$-term into baryons and antibaryons, the subsequent
annihilation of matter and antimatter would result in a homogeneous
gamma-ray background in the universe \cite{FR}. A decay of
the cosmological term
directly into radiation could be probed by the microwave background
anisotropies and the cosmological nucleosyntethis.
Supposing that the cosmological constant is decaying via particles,
the present rate of the
particle creation
(annihilation)
$$n=\frac{1}{R_0^3}\frac{d(\rho R^3)}{dt}|_0,$$
(where the subscript '0' denotes the present value of the corresponding
quality) in the considered model is $2\rho H$ less
than in Steady state cosmology ($3\rho H$).\
We remark that the  free energy of the decaying $\lambda$ may cause also
other effects than the creation of particles (nucleons) or radiation.
It can be
stored,
e.g. in form of small vacuum excitations of the gravitation field (see
\cite{MM}).
The detailed discussion of this topic would exceed the scope of this
paper.

\section{Final remark}

Summing up, we can state:\\
(i) In previous sections we have shown
that in the basic dynamical
equation (17) the energy density does not explicitly appear only the
density parameter $\Omega_M$.
We note that the density parameter $\Omega_M$ as the ratio of $\rho_M$
and $\rho_{crit}$ may be {\it finite} although both quantities are
infinite.\\
(ii) In the recently popular $\Lambda$CDM cosmological model, which
consists of a mixture of vacuum energy and cold dark matter, a serious
problem exists called in \cite{ST} as the cosmic  coincidence problem.
Since the vacuum energy density is constant over time and the matter
density decreases as the universe expands it appears that their ratio
must be set to immense small value ($\approx 10^{-120}$) in the early
universe
in order for the two densities to nearly coincide today, some billions
years later. No coincidence problem exists in our model of the universe
because $\Lambda_A$ here is functionally connected with
$\Omega_M$ in such a
way that this ratio in the matter dominated epoch does not vary over
time.\\
(iii) In the radiation dominated epoch $w=1/3$ and, according to
Eq.(8), $T=0$. The evolution dynamics
in this epoch runs so as if $\lambda =0$. Let us remark that in the
string-dominated universe ($w=2/3$) the cosmological constant becomes
negative!

The relation between $w_Q$ and $\kappa$ is \cite{MAL}
$$w_Q=-\frac{\kappa}{1+\kappa}.$$
 New measurement required $w_Q \leq -0.7$ \cite{GA}.
Inserting $\kappa$ for could dark matter $Q_M \approx 0.3$ we have
$w_Q \leq -0.7$. It is noteworthy, that in the limiting case when
$Q_M \rightarrow 0$ $\kappa \rightarrow \infty$, i.e. an almost empty
spacetime behaves similar as a space-time with the static cosmological
constant. In \cite{MAL} there are graphs of the angular-diameter
distance on redshift for the Friedmann cosmologies with selected value
of $Q_M$ and  $\kappa$.
 
In conclusion, the cosmological parameters of our cosmological
models are comfort with
the recent observational data of the flat and acceleration universe.
The described universe is
leigh-weight, is strictly flat, is accelerating, is old enough and is
causally
connected. One can speculate about the linear functional dependence
of $\Lambda$ on $T.$ The simplest hypothesis seems to be that $T$
is source of an unknown classical field whose quintessence energy density is
proportional to $\Omega_M$.

\end{document}